\documentclass[a4paper, 12pt]{article}

\usepackage {amssymb, amsmath, bbm, graphicx}
\def\b{\begin{eqnarray}}
\def\e{\end{eqnarray}}
\def\bn{\begin{eqnarray*}}
\def\en{\end{eqnarray*}}
\def\>{\rangle}
\def\<{\langle}
\title{Elementary Model of\\ Constraint Quantization with an Anomaly} 
\author{J. Scott Little\footnote{little@phys.ufl.edu}\hskip.4cm and John R. Klauder\footnote{klauder@phys.ufl.edu}~\footnote{Also Department of Mathematics}\\
Department of Physics\\
University of Florida\\
Gainesville, FL  32611}
\begin{document}
\date{}
\maketitle
  \begin{abstract}
 Quantum gravity is made more difficult in part by its constraint structure.  The constraints are classically first-class; however, upon quantization they become partially second-class.  To study such behavior, we focus on a simple problem with finitely many degrees of freedom and demonstrate how the projection operator formalism for dealing with quantum constraints is well suited to this type of example. 
  \end{abstract}
\section{Introduction}
It is well known that  the constraint algebra for classical gravity is first-class; however, following the standard Dirac prescription for the quantization of constraints, the resultant quantum constraint 
algebra is partially second-class \cite{dirac1}.  Since the Dirac procedure is better adapted to quantum first-class systems, the partial second-class character 
causes difficulties when discussing this problem.  Another procedure for constraint quantization is 
known as the projection operator formalism; for an overview of this method see \cite{klauder}.  One advantage this
method has over the Dirac procedure is that it treats first- and second-class quantum constraints on an equal
footing.  The premise behind this method is that one first quantizes a classical system and then reduces the 
original Hilbert space via a suitable projection operator to a subspace in which the constraints are satisfied to an appropriate level of accuracy; see \cite{klauder} for a more complete discussion.\footnote{As a note of interest, one could consider the case of the simplest example of a classical second-class constraint, namely, $p=q=0$.  When mapped into a corresponding quantum system, the constraints are satisfied by a projection operator onto the ground state of an harmonic oscillator with $\omega =1$, $\mathbbm{E}(P^2+Q^2\le\hbar) =|0\>\<0|$ \cite{klauder}.}

Explicitly, we recall that the constraint algebra for classical gravity is given by
\b
\{H_a(x),H_b(y)\} &=& \delta_{,a}(x,y)H_b(x)-\delta_{,b}(x,y)H_a(x),\label{eq1}\\
\{H_a(x),H(y)\} &=& \delta_{,a}(x,y)H(x),\label{eq2}\\ 
\{H(x),H(y)\} &=& \delta_{,a}(x,y)g^{ab}(x)H_b(x),\label{eq3}
\e
where $H$ is the Hamiltonian constraint and $H_a$ are the diffeomorphism constraints.  While (\ref{eq1}) and (\ref{eq2}) maintain a first-class structure upon quantization, (\ref{eq3}) morphs into a partially second-class constraint due to the metric factor; see \cite{anderson} for a further discussion.  It is this particular behavior of gravity that we mimic with a simple, few-degrees of freedom quantum mechanical system. 

\section{Classical Formulation}
The initial problem under consideration is based on using the three components of angular momentum, $j_i$, 
$i \in \{1,\,2,\,3\}$, as constraints. As a modification of this problem, we multiply each component of the angular momentum by a suitable, non-vanishing variable. It is convenient to
describe both of these examples in parallel.

\subsection{Basic Constraints} 
 We discuss a system of constraints that force the angular momentum $j_i$, $i\in \{1,\,2,\,3\}$, to vanish. With the angular momentum $j_i\equiv \epsilon_{ijk}q^jp_k$ (summation implied), the action integral we choose is given by 
\begin{equation}
I_1 = \int\left(p_a\dot{q}^a - \lambda^b j_b\right)dt,
\end{equation}
where $\lambda^b$ denotes the Lagrange multipliers to enforce the constraints.  Note that the Hamiltonian $H(p,q)$ is identically zero in this example. This form has been chosen for simplicity so we can focus directly on the issues surrounding
the constraints. 

{}From the definition of the $j_i$'s, one can immediately determine the Poisson algebra, given as usual by
\b
\{j_i,j_j\} = \epsilon_{ijk} j_k\;.
\e
Since this bracket yields a Lie algebra, our system is clearly a closed first-class constraint system \cite{henneaux}.
 
\subsection {Modified Constraints}
The action for our choice of the modified system is very similar in form, i.e.,
\begin{equation}
I_2 = \int\left(p_a\dot{q}^a - \lambda^b l_b\right)dt\; \label{eq6},
\end{equation}
where the essential change resides in the definition of the variables $l_i$.
For some smooth, non-vanishing function, $f$, we define (note: $q_1 \equiv q^1$, etc.) 
\begin{equation}
l_i \equiv f(p_1,p_2,q_1,q_2)\,j_i, 
\end{equation}
for all $i$,  and choose for further study the particular example for which
\begin{equation}
  f(p_1,p_2,q_1,q_2)\equiv \alpha +(\beta/{\tilde\hbar})(p_1^2+q_1^2)\ +(\gamma/{\tilde\hbar})(p_2^2+q_2^2). 
\end{equation}
The symbol ${\tilde\hbar}$ is a fixed constant equal in value to the physical
value of Planck's constant $\hbar$, namely $1.06\times10^{-27}$ erg-sec.
When the classical limit is called for, and thus Planck's constant $\hbar
\rightarrow0$, we emphasize that ${\tilde\hbar}$ retains its original
numerical value. We have chosen such a small divisor to emphasize the
quantum corrections; different divisors can be considered by rescaling
$\beta$ and $\gamma$. 

Since $\hbar$ dependence will play a role in our analysis, we do not set $\hbar =1$. In our choice of units, both $p$ and $q$ have the dimensions of $\sqrt{\hbar}$. Therefore, the division of $\beta$ and $\gamma$ by ${\tilde\hbar}$ has been chosen so that $\alpha$, $\beta$, and $\gamma$ can all be dimensionless.  We further suppose that $\alpha+\beta+\gamma=1$ and $0 < \alpha\le 1$; hence, $0 < \alpha \le f$.  Due to an evident $\beta \leftrightarrow \gamma$ symmetry it suffices hereafter to consider $0\le \gamma < \beta$.  (The special case of $\gamma=\beta$ is considered later.) 

We recognize, in this simple case, that we could absorb the factor $f$ by a redefinition of the Lagrange multipliers in (\ref{eq6}).  In more complicated systems (e.g., gravity) this simplification is either very difficult or perhaps impossible.  Therefore, as a further analog, we retain $f$ as a part of $l_a$.  A straightforward analysis leads to
\bn
\{l_i , l_j \} &=& \{fj_i, fj_j \} \\
               &=& f^2\{j_i , j_j \} + f\{j_i,f\}j_j+f\{f,j_j\}j_i + 
\{f,f\}j_ij_j \\
               &=& f \epsilon_{ijk}l_k + \{j_i,f\}l_j +\{f,j_j\}l_i\\
               &=&  f\epsilon_{ijk}l_k + 
\epsilon_{iab}[-q^a \partial f/\partial q^b +p_b \partial f/\partial p^a]l_j\\
               & &  -\quad \epsilon_{jab}[-q^a\partial f/\partial q^b +
p_b\partial f/\partial p^a]l_i\;. 
\en
Since $f>0$, our modified set of constraints is classified as open, first-class.

For the particular example of interest,
\begin{eqnarray}
\nonumber \{l_i,l_j\}&=&[\alpha +(\beta/{\tilde\hbar})(p_1^2+q_1^2)+(\gamma/{\tilde\hbar})(p_2^2+q_2^2)]\,\epsilon_{ijk}\,l_k\\
\nonumber & & -2(\beta/{\tilde\hbar})\,\epsilon_{ia1}(q^aq_1+p_ap_1)\,l_j+2(\beta/{\tilde\hbar})
\,\epsilon_{ja1}\,(q^aq_1+p^ap_1)\,l_i \\
 & & -2(\gamma/{\tilde\hbar})\,\epsilon_{ia2}(q^aq_2+p_ap_2)\,l_j+2(\gamma/{\tilde\hbar})
\,\epsilon_{ja2}\,(q^aq_2+p_ap_2)\,l_i,  
\end{eqnarray} \\
which is manifestly an open first-class constraint algebra when $\beta >0$. The equivalence of the two 
models is evident since $j_k=0$ implies that $l_k=0$, for all $k$, and
vice versa. Observe, as $\beta\rightarrow0$, and thus $\alpha\rightarrow1$, 
the $l$-constraints pass smoothly to the $j$-constraints. 
   
\section{Quantization}
We now proceed by canonically quantizing our model; a path integral quantization is discussed in Section 4.   Initially, let us  assume our chosen canonical coordinates are Cartesian coordinates suitable for quantization \cite{dirac2}, and thus we promote the variables  $(p_i, q^j)$ to a set
of irreducible, self-adjoint  operators $(P_i,Q^j)$ that obey the standard Heisenberg commutation relation,
  \b
[Q^i,P_j] = i \hbar \delta^i_j.
  \e
Following the standard Dirac procedure of constraint quantization, our constraints are chosen as suitable self-adjoint functions of the basic operators, namely
\bn
j_i &\mapsto& J_i = \epsilon_{ijk}Q_j P_k \\
l_i &\mapsto& L_i = \alpha J_i +(\beta/2\hbar)\,[(P_1^2+Q_1^2)\,J_i +J_i\,(P_1^2+Q_1^2)]\\
& & \hskip.6cm +\quad (\gamma/2\hbar)\,[(P_2^2+Q_2^2)\,J_i +J_i\,(P_2^2+Q_2^2)],  \en
where $\alpha +\beta +\gamma =1$. In dealing with the quantum theory, we drop the distinction between
${\tilde\hbar}$ and $\hbar$. 
A simple calculation shows that any other choice of factor-ordering to define $L_i$ is 
equivalent to the chosen form. In particular, normal ordering of the 
expressions is equivalent to a redefinition of the parameters $\alpha$, 
$\beta$, and $\gamma$. 

\subsection {Quantum Classification}
As in the classical case we can determine the classification of the $J_i$'s  based on their commutator structure as given by
\begin {eqnarray}
[J_i,J_j] = i \hbar \epsilon_{ijk} J_k \;.
\end {eqnarray}
This identity, which follows the classical Poisson bracket, implies 
that the $J_i$'s form a quantum system of closed, first-class constraints.

On the other hand, if we look at the commutators for the $L_i$'s, as shown below, we find a different outcome altogether. In what follows, we have introduced 
\b A\equiv (\alpha/2) + (\beta/2\hbar)\,(P_1^2+Q_1^2) + (\gamma/2\hbar)\,(P_2^2+Q_2^2),\e
 which means that
  \b
 L_j\equiv A\,J_j+J_j\,A . 
\e
Observe that the operator $A>0$. However, unlike the case of the classical constraints, the equation $J_l|\phi\>=0$, for some $l$, does {\it not} generally imply
that $L_l|\phi\>=0$, whenever $\beta>0$. 
 The commutator of $L_i$ and $L_j$ reads
\begin {eqnarray*}
[L_i, L_j] &=& i\hbar \epsilon_{ijk}AJ_kA +(J_iAJ_j -J_jAJ_i)A + A(J_iAJ_j -J_jAJ_i)\\
 & & \hskip.5cm+ (J_iA^2J_j -J_jA^2J_i).\\
\end {eqnarray*}

Unlike the $J_i$'s, the $L_i$'s are no longer first-class constraints, but are partially second class, a consequence of the open first-class nature
of their classical system. The second-class nature of the $L_i$'s is exhibited  below.

\subsection {Restricted Quantum Problem}
A complete analysis of the $J$- and $L$-algebras is sufficiently complicated that we seek a simplification. To this end, let us introduce conventional annihilation and creation operators given by 
   \b a_j=(Q_j+iP_j)/\sqrt{2\hbar}\;,  \\
      a^\dagger_j=(Q_j-iP_j)/\sqrt{2\hbar}\;.  \e
If we define
  \b N=a^\dagger_1a_1+a^\dagger_2a_2+a^\dagger_3a_3 \e
as the total number operator, it is evident that
  \b  [J_j,N]=0\;,\hskip1cm [L_j,N]=0\;,  \e
for all $j$, and thus both sets $\{J_i\}$ and $\{L_i\}$ are {\it number conserving}. 
This conservation implies that we can study the fulfillment of both sets of constraints in each of the  
number-operator subspaces independently of one another. We observe
that the subspace for which $N=0$ consists of just a single state, and this
state is an eigenvector of each $J_i$ as well as each $L_i$, $i\in\{1,2,3\}$, 
all with eigenvalue zero. In the 
interest of simplicity in this 
paper, we restrict our attention to the lowest nontrivial 
subspace in which
the constraints $J_i=0$ are satisfied on a non-vanishing subspace. In particular, we confine our attention to a subspace of the entire Hilbert space 
corresponding to an eigenvalue of the total number operator of two.  Note
that the subspace of interest is six-dimensional and is spanned by the six vectors given by the two representatives
  \b  |1,1,0\>&=&a^\dagger_1\,a^\dagger_2\,|0\>\;,  \hskip.5cm{\rm etc.,}\\
      |2,0,0\>&=&(1/\sqrt{2})\,(a^\dagger_1)^2\,|0\>\;,\hskip.5cm{\rm etc.}\;, \e
where as usual $|0\>\, (=|0,0,0\>)$ denotes the no particle state for which $a_j|0\>=0$
for all $j$. 

\subsection{Construction of the Projection Operators}
Having chosen a particular subspace of the Hilbert space on which to focus our attention, we now have the information needed to construct the projection operators.  We do so by first determining all possible eigenvectors and eigenvalues of the equation
\begin{equation}
\Sigma^3_{i=1} \Phi_i^2\,|\psi\rangle = \nu |\psi\rangle\;,
\end{equation}
where $\Phi_i = J_i$ or $L_i$, and $\nu\ge0$.
We also consider the equation
\begin{equation}
\Phi_3 \,|\psi\rangle = \eta|\psi \rangle\;, 
\end{equation}
where $\Phi_3 = J_3$ or $L_3$, and $\eta$ is real.

\subsection{$J_i$ Considerations}
We begin with our quantum first class system.  The well-known eigenvectors and eigenvalues for the $N=2$ subspace are presented in Table 1 (modulo normalization factors for the eigenstates).\\
\begin{center}Table 1:  $J_3$ and $J^2$ Eigenstates and Eigenvalues \end{center}
\vskip.4cm
\begin{tabular}{|c|c|c|}
\hline
$Eigenstates$&$\mbox{$J_3$ Eigenvalues}$&$\mbox{$J^2$ Eigenvalues}$\\
\hline
\hline
$\big|1,1,0\rangle - \frac{i}{\sqrt{2}}\big|2,0,0\rangle + \frac{i}{\sqrt{2}}\big|0,2,0\rangle$ 
&$2\hbar$&$6\hbar^2$\\
\hline
$\big|1,0,1\rangle +i\big|0,1,1\rangle$&$\hbar$&$6\hbar^2$\\
\hline
$\big|2,0,0\rangle +\big|0,2,0\rangle -2\big|0,0,2\rangle $ & 0 &$6\hbar^2$\\
\hline
$\big|1,0,1\rangle - i \big|0,1,1\rangle $&$-\hbar$&$6\hbar^2$\\
\hline
$\big|1,1,0\rangle + \frac{i}{\sqrt{2}}\big|2,0,0\rangle - \frac{i}{\sqrt{2}}\big|0,2,0\rangle$ 
&$-2\hbar$&$6\hbar^2$\\
\hline\hline
$\big|2,0,0\rangle +\big|0,2,0\rangle + \big|0,0,2\rangle $ &$ 0$ &$ 0$\\
\hline
\end{tabular}
\vskip.4cm
As expected, the operators $J_3$ and $J^2\equiv\Sigma J_i^2$ share a common basis, since they can be simultaneously diagonalized.  Based on this information, we choose our projection operator as 
\begin{eqnarray*}
\mathbbm{E}_J &\equiv &\mathbbm{E}(\Sigma_{i=1}^3 J_i^2 \le \hbar^2) \\ 
&=&
{\textstyle\frac{1}{3}}\,(\,|2,0,0\rangle + |0,2,0\rangle +|0,0,2\rangle\,)(\,\langle 2,0,0| + \langle 0,2,0| + \langle 0,0,2|\,)\\
&\equiv & |O_J\rangle \langle O_J|\;.
\end{eqnarray*}
Note that the factor $\hbar^2$ in the argument of the projection operator
has been selected to exclude any integer spin value other than zero; clearly, other
values for that parameter (less than $6 \hbar^2$) would do equally well.

We can now see how the projection operator determines the one-dimensional 
physical Hilbert space ${\cal H}_{P}$ within the six-dimensional original Hilbert space (confining attention to the two-particle subspace) as follows.  Consider a general vector $|\psi\rangle_2 \in \mathcal{H} _2 $ (the two-particle subspace). Then
\begin {eqnarray}
\mathbbm{E}_J |\psi \rangle _2& = & |O_J\rangle \langle O_J|\psi \rangle_2 =|\psi\>_{2P} \in \mathcal{H} _{2P},\\ 
J_i \mathbbm{E}_J |\psi \rangle_2 & = &J_i |O_J\rangle \langle O_J|\psi \rangle_2 = 0 
\in{\cal H}_{2P},\hskip.7cm i=1,2,3\;. 
\end{eqnarray}
This result implies that our constraint condition ($J_j|\psi\rangle_{2P} = 0$ ,\quad for all $j$) is satisfied by every vector in the physical Hilbert space.  Consequently, we have successfully quantized our closed first-class constraints, in this restricted Hilbert space.
It is apparent for this example that a similar story would apply in all even-particle subspaces of the original Hilbert space.

\subsection {$L_i$ Considerations}
We next follow a similar procedure for the $L_i$ operators. To further simplify matters, we choose $\gamma \equiv \beta/2$. The eigenvectors and eigenvalues of $L_3$ and $L^2\equiv\Sigma L_i^2$ are presented in Tables 2 and 3 (modulo normalization factors for the eigenstates).\\
\begin{center} Table 2:  $L_3$ Eigenstates and Eigenvalues
\vskip.4cm
\begin{tabular}{|c||c|}
\hline
$L _3 \mbox{ Eigenstates}$&$L_3 \mbox{ Eigenvalues}$\\
\hline
\hline
$\big|1,1,0\rangle - \imath a\big|2,0,0\rangle 
+ \imath a'\big|0,2,0\rangle$ 
&$\hbar\sqrt{4+24\beta+37\beta^2}$\\
\hline
$\big|1,0,1\rangle +\imath \big|0,1,1\rangle$ & $\hbar\left(1+3/2 \beta\right)$\\
\hline
$\frac{2+5\beta}{2 +7\beta}\big|2,0,0\rangle +\big|0,2,0\rangle -\frac{2\left(4+24\beta +37\beta^2\right)}{\left(2+ 7\beta\right)^2}\big|0,0,2\rangle $ & 0\\
\hline
$\big|1,0,1\rangle -\imath \big|0,1,1\rangle$&$-\hbar\left(1 +3/2 \beta\right)$\\
\hline
$\big|1,1,0\rangle + \imath a\big|2,0,0\rangle 
- \imath a'\big|0,2,0\rangle$ 
&
$-\hbar\sqrt{4+24\beta +37\beta^2}$\\
\hline
\hline
$\frac{2+5\beta}{2+7\beta}\big|2,0,0\rangle +\big|0,2,0\rangle + \big|0,0,2\rangle $ & 0\\
\hline
\end{tabular}
\end{center}
Here $a$ and $a'$ are defined as,
\begin{eqnarray*}
a  &=& \frac{2+7\beta}{\sqrt{8+48\beta+74\beta^2}},\\
a' &=& \frac{2+5\beta}{\sqrt{8+48\beta+74\beta^2}}.\\
\end{eqnarray*}
Observe how these $L_3$ eigenstates and eigenvalues all pass smoothly to the appropriate $J_3$ eigenstates and eigenvalues as $\beta \rightarrow 0$.\\

We now turn our attention to the eigenvectors and eigenvalues of $L^2$, and in this discussion we offer exact and approximate statements.  The first three entries in Table 3, below, associated with the states $|1,1,0\rangle$, etc., are {\it exact} in both the eigenvectors and eigenvalues. The {\it{product}} of the last three eigenvalues, associated with the three states $|2,0,0\rangle$, etc., is given {\it{exactly}} by the expression $288\,\beta^2\gamma^2(\beta -\gamma)^2$.  This relation implies that (at least) one eigenvalue is zero if $\beta=0$, or $\gamma=0$, or $\beta=\gamma$.  Thus we confine attention to the range $0\le\gamma<\beta$, and focus on the case where $\gamma=\beta/2$. Unfortunately, the exact forms of the last three eigenvectors and eigenvalues are exceedingly complicated functions of $\beta$ and provide little insight into their small $\beta$ behavior.  To present this information, we use their Taylor series to first non-zero order in $\beta$.  The exact nature of these three eigenvalues can be observed in Figs. 1 and 2.

\begin {center}Table 3:  $L^2$ Eigenstates and Eigenvalues 
\vskip.3cm
\begin{tabular}{|c|c|}
\hline
$\Sigma L_i^2 \mbox{ Eigenstates}$&$\Sigma L_i^2 \mbox{ Eigenvalues}$\\
\hline
\hline
$\big|0,1,1\rangle $& $\hbar^2\left(6+15\beta + 45\beta^2/4\right)$\\
\hline
$\big|1,0,1\rangle $ & $\hbar^2\left(6+ 24\beta +57\beta^2/2\right)$\\
\hline
$\big|1,1,0\rangle $ & $\hbar^2\left(6+33\beta +189\beta^2/4\right)$\\
\hline
$b\big|2,0,0\rangle +b'\big|0,2,0\rangle +\big|0,0,2\rangle$& $\hbar^2\left(6+6(4+\sqrt{3})\beta +O(\beta^2)\right)$\\
\hline
 $c\big|2,0,0\rangle +c'\big|0,2,0\rangle +\big|0,0,2\rangle$ &$\hbar^2\left(6+6(4-\sqrt{3})\beta +O(\beta^2)\right)$\\
\hline
\hline
$d\big|2,0,0\rangle +d'\big|0,2,0\rangle +\big|0,0,2\rangle$ & $\hbar^2\left(\beta^6/2 +O(\beta^7) \right)$\\
\hline
\end{tabular}
\end{center}
\vskip.4cm
Here $b$, $b'$, $c$, $c'$, $d$ and $d'$ are approximated by
\begin {eqnarray*}
b  &=& (-2-\sqrt{3})-\frac{7}{4}(2+\sqrt{3})\beta +O(\beta^2),\\
b' &=&  (1+\sqrt{3})+(\frac{1}{2}+\frac{3\sqrt{3}}{4})\beta+ O(\beta^2),\\
c  &=& (-2+\sqrt{3})-\frac{7}{4}(-2+\sqrt{3})\beta +O(\beta^2),\\
c' &=& (1-\sqrt{3})+(\frac{1}{2}-\frac{3\sqrt{3}}{4})\beta +O(\beta^2),\\
d  &=& 1- 2\beta + O(\beta^2),\\
d' &=& 1-\beta+O(\beta^2)\;.\\
\end {eqnarray*}
Clearly, as $\beta \rightarrow 0$, and thus $\alpha \rightarrow 1$, we recover all the properties of $J_3$ and $\Sigma J_i^2$.\\
\begin{center}
\includegraphics[scale=0.5]{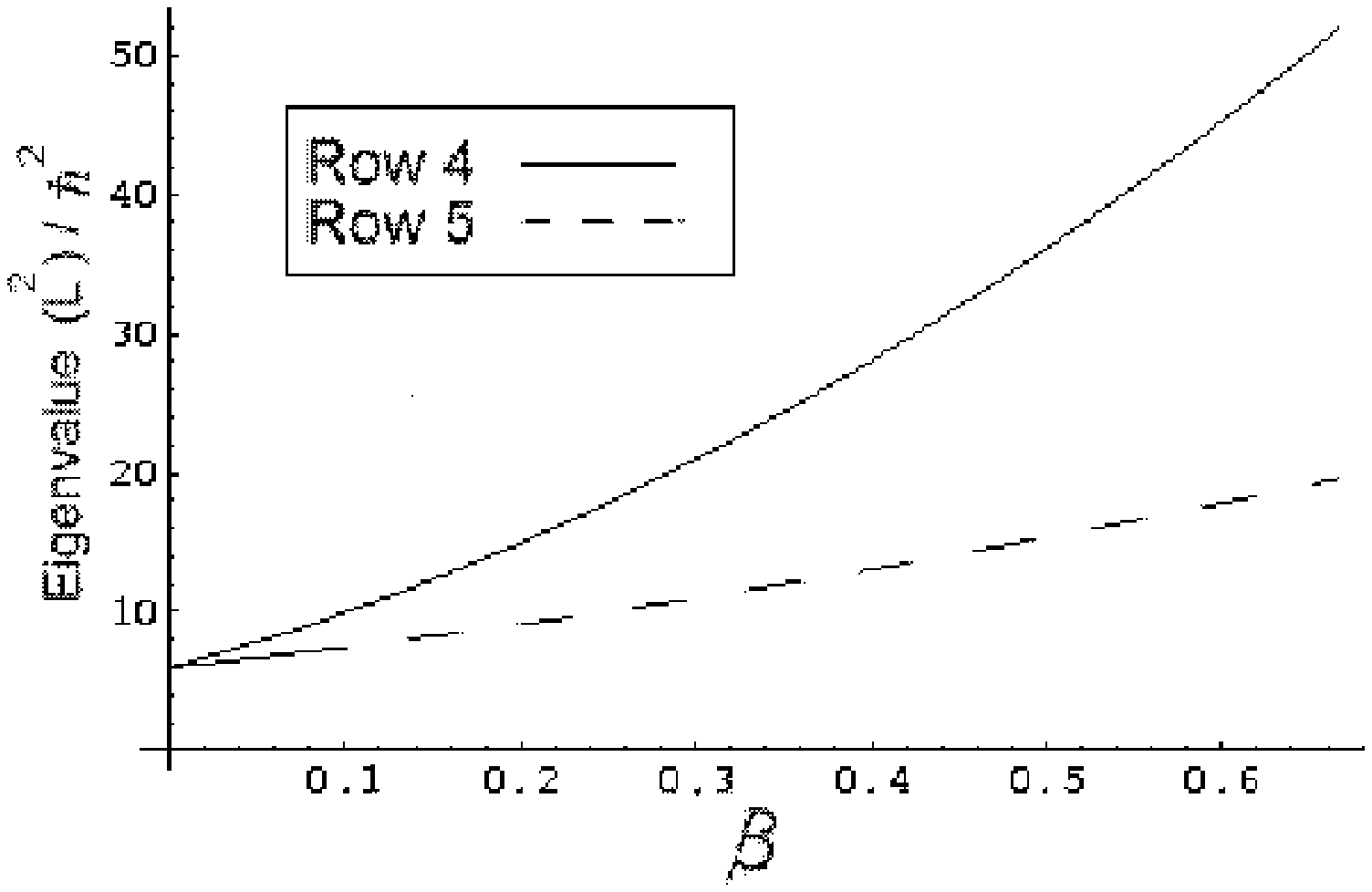}\\
Fig.1 Exact $L^2$ Eigenvalues of Row 4 and Row 5, Table 3, as functions of $\beta$\\
  \includegraphics[scale=0.5]{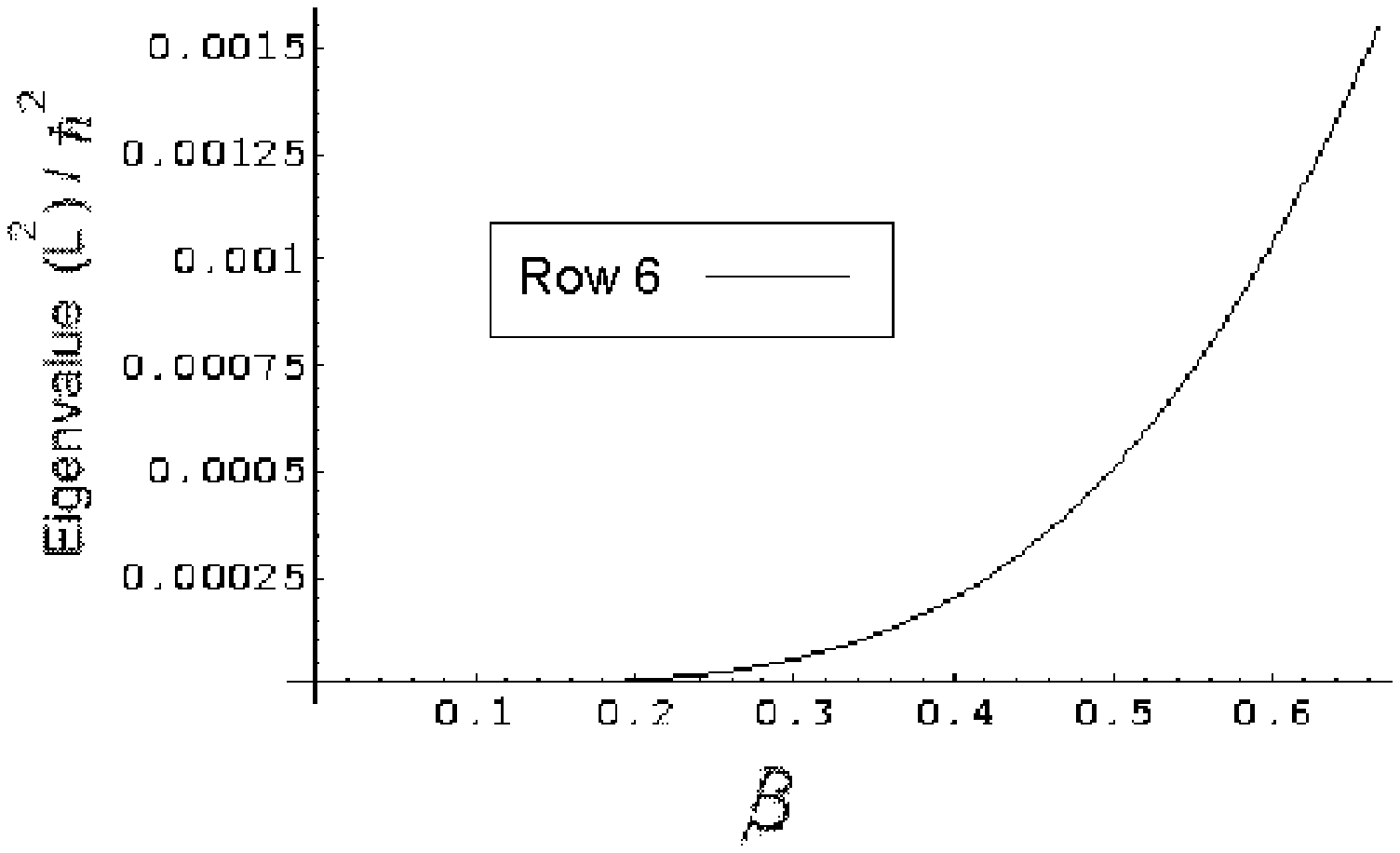}\\
Fig.2 Exact $L^2$ Eigenvalue of Row 6, Table 3, as a function of $\beta$ \\
\end{center}
Unlike $J_3$ and $\Sigma J_i^2$, the operators $L_3$ and $\Sigma L_i^2$ do {\it not} share a common basis.   An important feature of the $ L^2$ operator is the {\it absence} of zero in its spectrum when $\beta > 0$. This is a clear reflection of the partial second-class nature of the $L_i$ constraints.  However, the spectrum of $\Sigma L_i^2$  admits a lowest possible eigenvalue, and from that state (or states) we construct our projection operator. To ensure that
$\alpha>0$, when $\gamma=\beta/2$, it is necessary to restrict $\beta$
so that $0\le\beta<2/3$. It follows that the lowest eigenvalue of $\Sigma L_i^2$ is non-degenerate and is represented by the last entry in Table 3 (row 6). 
Thus, in this case, we are led to adopt the projection operator given by
\begin {eqnarray}
\nonumber \hskip-.3cm \mathbbm{E}_L&\equiv&\mathbbm{E}(\Sigma_{i=1}^3 L_i^2 \le \delta (\hbar)^2) \\
\nonumber\hskip1cm&=&|M|^2(\,d|2,0,0\rangle + d'|0,2,0\rangle +|0,0,2\rangle\,)
(\,d\langle 2,0,0| +d'\langle 0,2,0|+\langle 0,0,2|\,)\nonumber\\
&\equiv& |O_L\rangle  \langle O_L|\;,
\end {eqnarray}
where $M = 1/\sqrt{d^2+d'^2+1}$ is a normalization factor. In this expression we may set 
 \b \delta (\hbar)^2 = \beta \hbar^2 \e
which lies between the lowest and next to lowest eigenvalues of $\Sigma L_i^2$ when $0\le\beta<2/3$ (i.e.. ${\rm Eigenvalue}(L^2)_{\mbox{ row 6}} \le \delta(\hbar)^2 $).  Other choices for $\delta (\hbar)^2$ are possible as long as the resultant $\mathbbm{E}_L$ is the same.   
Again by taking an arbitrary vector $|\psi \rangle_2 \in \mathcal{H} _2 $, we can project that vector into the two-dimensional physical Hilbert space  by acting with $\mathbbm{E}_L$ on the vector. In symbols, we have 
\begin{equation}
\mathbbm{E}_L\,|\psi\>_2\equiv|\psi \rangle _{2P} = | O _L\rangle   \langle O_L|\psi\rangle_2 \in \mathcal{H} _{2P}
\end{equation}

Unlike our first class constraint system, the constraint equation $L_i|\psi\rangle_{2 P} \ne 0$  for all $i\in \{1,2,3\}$, unless $|\psi\rangle_{2P} =0$. 
However, this behavior is a general characteristic of quantum second-class 
constraints.

 With the foregoing discussion, we have successfully dealt with a quantum second-class constraint in the context of our restricted Hilbert space. An analogous discussion would take place in higher-order, even-particle subspaces. 

\section{Path Integral Approach}

In Section 3, we discussed the canonical quantization of the model, however, this is not the only procedure to quantize the system.  We now focus our attention on a path integral approach.  Before discussing the projection operator in this language, let us first show how the Faddeev method \cite{faddeev} would treat this example.  The formal starting relation
\begin{equation*}
 \int e^{(i/\hbar) \int[p_a\dot{q}^a  -\lambda^b fj_b]dt}{\cal D}p {\cal D}q {\cal D}\lambda, \\
\end{equation*}
is replaced with the gauge-fixed expression
\begin{equation}
 \int e^{(i/\hbar)\int p_a \dot{q}^a dt}\Pi_b \Pi_c \delta\{\chi^b\}\delta\{fj_b\} \det \{\chi^b, fj_c\}{\cal D}p {\cal D}q,\label{eq27}
\end{equation}
where $\chi^b (p,q)$ is some appropriate gauge choice.  A simple identity leads to 
\b
\int e^{(i/\hbar) \int p_a \dot{q}^a dt}\Pi_b \Pi_c \delta\{\chi^b\}\frac{\delta\{j_b\}}{\Pi_t f^3}\det(\{\chi^b,f\}j_c+\{\chi^b,j_c\}f){\cal D}p {\cal D}q.
\e
The first term in the determinant is zero by the $\delta$ functional of the $j$'s.  The second term is a $3 \times 3$ matrix multiplied by a scalar $f$, and therefore becomes.
\b
\int e^{(i/\hbar) \int p_a\dot{q}^a dt}\Pi_b \Pi_c \delta\{\chi^b\}\frac{\delta\{j_b\}}{\Pi_t f^3}\Pi_t f^3 \det\{\chi^b,j_c\} {\cal D}p {\cal D}q.
\e
We observe that all  the factors of $f$ completely cancel.
As one can see the Faddeev method is insensitive to the definition of $f$, only that it be non-zero.  Hence, this method considers the $j$'s and $l$'s as identical constraints.  This is not surprising since using this method we are satisfying the constraints classically, and as shown in Section 2, the constraints behave the same in the classical regime.  

It is often stated that the results of (\ref{eq27}) are correct up to terms of order $\hbar$.  While this may be true, the goal of quantization is to obtain correct $\hbar$ dependence.  Otherwise Bohr-Sommerfeld \cite{shankar} would be an adequate quantization procedure.  Based on the discussion of Section 3 we can acquire the correct $\hbar$ dependence in the present case.

A coherent state path integral can also be used to calculate the matrix elements of the projector as shown in previous works \cite{klauder}.  Let us begin with a preliminary equation, namely,   

\begin{eqnarray*}
\< \vec z \,''|Te^{-(i/\hbar)\int \lambda^a J_a dt}| \vec z \,'\> & & \\
&& \hskip-1cm ={\cal M} \int \exp\{(i/\hbar)\int((p_a\dot{q}^a-q^a\dot{p}_a)/2 - \lambda^a j_a)dt\}{\cal D}p {\cal D}q\\
&& \hskip-1cm =N'' N'\exp\{\vec z\,''^*\cdot e^{-(i/\hbar)\vec\theta \cdot \vec j}\vec z\,'\},\\ 
\end{eqnarray*}
where ${\cal M}$, $N''$, and $N'$ are normalization factors, $\vec z \equiv (\vec q +i \vec p)/\sqrt{2\hbar}$, $\vec j$ is a $3\times3$ matrix representation of the rotation algebra, $T$ denotes time ordering, and $\vec \theta$ is a suitable functional of $\{\lambda^a(\cdot)\}$.

Following \cite{klauder}, we could integrate over $\vec \lambda$ with respect to a suitable measure ${\it R}(\vec \lambda)$ to create the desired projection operator.  However, it is equivalent and simpler to proceed as follows,  
\begin{eqnarray*}
{\cal K}_J(\vec z \,'';\vec z') &\equiv& \<\vec z\,''|\mathbbm{E}(J^2\le\hbar^2)|\vec z \,'\> \\
 &=& \int\<\vec z \,''|e^{-(i/\hbar)\vec\theta \cdot \vec J}|\vec z\,'\> d\mu(\theta),
\end{eqnarray*}
 where $d\mu (\theta)$ is the normalized Haar measure of $SO(3)$. Consequently,\\
\begin{eqnarray}
\nonumber{\cal K}_J(\vec z \,'';\vec z \,') &=&(  N'' N'/2)\int \exp\{\sqrt{\vec z \,''^{*2}}\sqrt{ \vec z \,'^2}\cos \theta\}d \cos\theta\\
 &=&  N'' N'\, \frac{\sinh {\sqrt{\vec z \, ''^{*2}\vec z'^2}}}{\sqrt{{\vec z \,''^{*2} \vec z'^2}}}\\
 &=&  N'' N'[1 +\frac{\vec z \,''^{*2}\vec z \,'^2}{3!}+ \frac{(\vec z''^{*2}\vec z \,'^2)^2}{5!}+ \, ... \,]\label{eq31}\\
&=& \<\vec z''|0\>\<0|\vec z \,''\>+\<\vec z \,''|O_J\>\<O_J|\vec z \,'\>+...\, .
\end{eqnarray}
From (\ref{eq31}) we can deduce that the physical Hilbert space for every even particle sector is one-dimensional,  while in the odd particle sectors the physical Hilbert space is empty, as conjectured in Section 3.  

We can also construct the fundamental kernel for the modified case using the results of Section 3.  Specifically,.  
\b
{\cal K}_L(\vec z\,'';\vec z\,') &=& \<\vec z \,''|0\>\<0|\vec z \,'\>+\<\vec z \,''|O_L\>\<O_L|\vec z \,'\>+...\\
\nonumber &=&  N'' N'[1+\frac{(d {z_1''}^{*2}+d'{z_2''}^{*2} +{z_3''}^{*2})(d z_1'^2+d'z_2'^2 +z_3'^2)}{2!(d^2+d'^2+1)}+...\,].\\
\e
Although the full nature of this kernel is unknown at present, we expect that every higher even sector contained in ${\cal K}_L$ will also yield a one-dimensional physical Hilbert space, based on the observed nature of the $J$'s. 
\section{Summary}

Motivated by the quantization of gravity, we have considered a system that has classically first-class constraints; however, upon quantization it transforms partially to second class.  In the two-particle subspace and for all $\beta >0$, our modified system has the characteristic feature that for a suitably chosen $\delta (\hbar)^2$, which generally depends on $\beta$, we have
\begin{equation}
0<\, _{2P}\langle \psi|\Sigma_i L_i^2|\psi\rangle_{2P} \le 
\delta (\hbar)^2\, _{2P}\langle \psi|\psi\rangle_{2P}\;,  
\end{equation}
an attribute shared by all quantum second-class constraints.    
If we were to take the limit $\delta (\hbar) \rightarrow 0$, our projection operator would vanish, resulting in an empty physical Hilbert space. Thus, as part of the general theory of the projection operator method \cite{klauder}, we do {\it not} take such a limit. Nevertheless, we observe that all properties of the $L$-constraints pass smoothly to those of the $J$-constraints as $\beta \rightarrow 0$, 
including the dimensionality of the physical Hilbert space. 

The results we have obtained could be extended to all particle sectors that admit a lowest possible eigenvalue. One may also be interested in extending our procedure to models with additional degrees of freedom, and finally extending the arguments to include fields.
\section{Acknowledgments}
Wayne Bomstad is thanked for his comments and encouragement on this paper, and the suggestions of Jan Govaerts and Sergei Shabanov proved helpful.  JSL would also like to acknowledge the University of Florida's Alumni Fellowship Program for support.

\end{document}